\documentclass[preprint,3p,sort&compress]{elsarticle}
\usepackage{amsmath}	
\usepackage{amssymb}
\usepackage{color}
\usepackage{caption}
\usepackage{graphicx}
\usepackage{subcaption}

\newcommand{\rz}{{\mathrm{Z}}}
\newcommand{\MZ}{\ensuremath M_{\rz}}
\def\xiG{\xi_G}
\def\xiW{\xi_W}
\def\xiB{\xi_B}
\def\alu{\ensuremath{a_t}}
\def\ald{\ensuremath{a_b}}
\def\ale{\ensuremath{a_\tau}}
\def\lam{\ensuremath{\hat\lambda}}

\def\Alu{\ensuremath{A_t}}

\def\Lam{\ensuremath{\Lambda}}

\newcommand{\ala}{\ensuremath {a_1}}
\newcommand{\alb}{\ensuremath {a_2}}
\newcommand{\alc}{\ensuremath {a_s}}

\newcommand{\Alb}{\ensuremath {A_2}}
\newcommand{\Alc}{\ensuremath {A_s}}

\newcommand{\MS}{{\ensuremath{\overline{\mathrm{MS}}}}}
\def\z#1{{\zeta_{#1}}}
\def\LYukawa{\ensuremath{\mathcal{L}_{\mathrm{Yukawa}}}}

\DeclareMathOperator{\tr}{tr}

\newcommand{\NR}{\ensuremath {N_c}}

\newcommand{\NGen}{\ensuremath {n_G}}

\newcommand{\cA}{\ensuremath {C_A}}
\newcommand{\cR}{\ensuremath {C_F}}

\def \eps {\epsilon}

\journal{Physics Letters B}
\begin{document}
\begin{frontmatter}
\title{Yukawa coupling beta-functions in the Standard Model at three loops}
\author[a]{A.~V.~Bednyakov}
\ead{bednya@theor.jinr.ru}
\author[a]{A.~F.~Pikelner}
\ead{andrey.pikelner@cern.ch}
\author[b]{and V.~N.~Velizhanin}
\ead{velizh@thd.pnpi.spb.ru}

\address[a]{Joint Institute for Nuclear Research,\\
 141980 Dubna, Russia}
\address[b]{Theoretical Physics Department, Petersburg Nuclear Physics Institute,\\
  Orlova Roscha, Gatchina, 188300 St.~Petersburg, Russia}

\begin{abstract}
We present the results for three-loop beta-functions for Yukawa couplings of heavy Standard Model
	fermions calculated within the unbroken phase of the model.
The calculation is carried out with the help of the \texttt{MINCER} program
	in a general linear gauge, and the final result is independent of the gauge-fixing parameters.
In order to calculate three-point functions, we made use of infrared  rearrangement (IRR) trick.
Due to the chiral structure of the SM a careful treatment of loops with fermions is required to perform
the calculation.
It turns out that gauge anomaly cancellation in the SM allows us to obtain the result by means of the semi-naive treatment of $\gamma_5$.
\end{abstract}

\end{frontmatter}

The Yukawa couplings being the fundamental parameters of the Standard Model (SM) Lagrangian
	describe the interactions of quarks and leptons with the Higgs field.
Having in mind the discovery of the Higgs boson candidate~\cite{:2012gk,:2012gu}
	one can hope that some day the values of Yukawa couplings will be deduced directly
	from the experimental data (see, e.g., the discussion presented in Ref.~\cite{Peskin:2012we}).
In order to obtain a very precise SM prediction for the running Yukawa couplings at some high energy scale,
	one usually uses known masses of quarks and leptons~\cite{Beringer:1900zz} since
	it is the Yukawa interactions that give
	the fundamental fermions their masses after spontaneous electroweak symmetry breaking. 	
Due to the observed hierarchy of the SM fermion masses the corresponding values are usually defined
	at different scales, so one inevitably makes use of renormalization group equations (RGE) to
	connect these scales.
The so-called threshold (matching) corrections
	(see Refs.~\cite{Hempfling:1994ar},~\cite{Jegerlehner:2012kn} for the case of running masses and Yukawa couplings) are also very important for extractions of the running
	SM parameters defined in the minimal subtraction (\MS) scheme, in which counter-terms
	and beta-functions have a very simple polynomial structure.
It also should be mentioned that contrary to leptons, quarks are not observed as free particles, so
	the pole mass which is usually associated with the physical mass of a particle,
	although being a gauge-invariant and infrared finite quantity~\cite{Tarrach:1980up,Kronfeld:1998di}, 
	suffers from the so-called renormalon ambiguity~\cite{Beneke:1994sw,Bigi:1994em}.
This intrinsic uncertainty of the quark pole mass is estimated to be of the order of
	$\Lambda_{QCD}\simeq 200$~MeV.
For the top-quark one usually neglects this uncertainty since the corresponding mass $M_t = 175$~GeV
	is much bigger than  $\Lambda_{QCD}$.
Moreover, it is generally believed that due to the short lifetime the $t$-quark does not have time to
	hadronize, so the notion of the pole mass can be used in this case.
According to the recent studies of the relation between the~\MS-running mass of the top quark
	and the corresponding pole mass, electroweak corrections can become important and for the
	observed value of the Higgs boson mass can severely cancel the QCD contribution~\cite{Jegerlehner:2012kn}.
As a consequence, theoretical uncertainty in determination of the top-quark Yukawa coupling
	is reduced, thus calling for more precise determination of the corresponding RGEs.
For all the other quarks one usually uses the~\MS~masses defined initially in the context of QCD (see, e.g.,
Ref.~\cite{Chetyrkin:2000yt} and references therein).

The SM Higgs boson with $M_h = 125$~GeV decays predominately into the $b\bar b$ pairs.
In spite of the fact that this decay mode is very hard to observe due to the enormous QCD background
	it is obvious that the precise value of the corresponding Yukawa coupling is required
	to test whether the SM correctly describes Nature at the electroweak scale.
If one considers leptonic decays of the Higgs boson, the most promising decay channel is the tree-level
	decay to a tau-anti-tau pair.
These facts somehow motivate our study of the three-loop contribution to the corresponding Yukawa beta-functions.

Moreover, we would like to stress here that a separate study of the high-energy behavior of the SM can also be
	of great importance.
One can use RGE to find the scale where New Physics should enter the game, e.g.,
	to unify the interactions or stabilize the Higgs potential~\cite{Krasnikov:1978pu,Hung:1979dn,Politzer:1978ic,Bezrukov:2012sa,Degrassi:2012ry,Alekhin:2012py}.

One-
and two-loop results for SM beta functions have been known for quite a long time~\cite{Gross:1973id,Politzer:1973fx,Jones:1974mm,Tarasov:1976ef,Caswell:1974gg,Egorian:1978zx,Jones:1981we,Fischler:1981is,Machacek:1983tz,Machacek:1983fi,Luo:2002ti,Jack:1984vj,Gorishnii:1987ik,Arason:1991ic}
and are summarized in~\cite{Luo:2002ey}.

Not long ago full three-loop gauge coupling beta-functions 
	were calculated~\cite{Mihaila:2012fm,Bednyakov:2012rb}.
The beta-functions for the Higgs self-coupling and top Yukawa coupling were also considered at three loops~\cite{Chetyrkin:2012rz}.
However, in Ref.~\cite{Chetyrkin:2012rz}, all the electroweak couplings were neglected together with the Yukawa couplings of
	other SM fermions.
In this paper, we provide the full analytical result for the three-loop beta-functions
	of the strongest Yukawa couplings $y_t, y_b, y_\tau$ for the three heaviest SM fermions (top, bottom, and tau).  We take into account all the interactions of the SM.

Let us briefly define our notation.
The full Lagrangian of the unbroken SM which was used in this calculation is given
	in our previous paper~\cite{Bednyakov:2012rb}.
However, we do not keep the full flavor structure of Yukawa interactions but
	use the following simple Lagrangian which describes fermion-higgs interactions
	\begin{equation}
		\LYukawa = - y_t (\bar{Q} \Phi^c) t_R - y_b (\bar{Q}\Phi) b_R - y_\tau (\bar{L} \Phi) \tau_R
		+ \mathrm{h.c.}
		\label{eq:yukawa_lag}
	\end{equation}
with $Q=(t,b)_L$, and $L=(\nu_\tau, \tau)_L$ being SU(2) doublets of left-handed fermions
of the third generation, $u_R$, $t_R$, and $\tau_R$ are the corresponding right-handed
	counter parts\footnote{Here we assume that neutrinos are massless.}.
The Higgs doublet $\Phi$ with $Y_W = 1$ has the following decomposition in terms of the component fields:
\begin{equation}
	\Phi =
	\left(
	\begin{array}{c}
		\phi^+(x) \\ \frac{1}{\sqrt 2} \left( h + i \chi \right)
		\end{array}
	\right),
	\qquad
	\Phi^c = i\sigma^2 \Phi^\dagger =
	\left(
	\begin{array}{c}
		\frac{1}{\sqrt 2} \left( h - i \chi \right) \\
		-\phi^-
		\end{array}
	\right).
	\label{eq:Phi_def}
\end{equation}
Here a charge-conjugated Higgs doublet is introduced $\Phi^c$ with $Y_W=-1$.

For loop calculations it is convenient to define the following quantities:
\begin{eqnarray}
	 a_i  & = &  \left(\frac{5}{3} \frac{g_1^2}{16\pi^2}, \frac{g_2^2}{16\pi^2}, \frac{g_s^2}{16\pi^2},
	 \frac{y_t^2}{16\pi^2}, \frac{y_b^2}{16\pi^2}, \frac{y_\tau^2}{16\pi^2}, \frac{\lambda}{16\pi^2},\xiG, \xiW, \xiG \right),
	 \label{eq:coupl_notations}
\end{eqnarray}
 	where we use the SU(5) normalization of the U(1) gauge coupling $g_1$.
We also stress that the calculation is carried out in a general linear $R_\xi$ gauge, so the renormalization
of all three gauge-fixing parameters $\xiG,~\xiW$, and $\xiB$, given in Ref.~\cite{Bednyakov:2012rb}, is required.

The Yukawa beta-functions are extracted from the corresponding renormalization constants
which relate bare couplings to the renormalized ones in the~\MS-scheme.
The latter are found with the help of the following formulae:
\begin{equation}
	Z_{y_f} = \frac{Z_{ffh}}{\sqrt{Z_{f_L} Z_{f_R} Z_{h}}}
	\qquad\mathrm{or}\qquad
	Z_{y_f} = \frac{Z_{ff\chi}}{\sqrt{Z_{f_L} Z_{f_R} Z_{\chi}}},\qquad f = t,~b,~\tau,
	\label{eq:Z_y}
\end{equation}
	where $Z_{ffh}$ and $Z_{ff\chi}$ are the renormalization constants for the three-point vertices involving
	two fermions $f$ and the higgs $h$ or the would-be goldstone boson $\chi$, respectively.
The renormalization constants $Z_{f_L}$, $Z_{f_R}$ for left- and right-handed fermions, and
	$Z_{\chi}$ and $Z_{h}$ for the neutral components of the Higgs doublet are obtained
	from the corresponding self-energy diagrams.

	In order to extract a three-loop contribution to the considered renormalization constants, 
	it is sufficient to know the two-loop results for the gauge and Yukawa couplings and the one-loop expression
	for the Higgs self-interaction.

The relation between the bare and renormalized parameters can be written in the following way
\begin{equation}
	a_{k,\mathrm{Bare}}\mu^{-2\rho_k\epsilon} = Z_{a_k} a_k(\mu)=a_k+\sum_{n=1}^\infty c_k^{(n)}\frac{1}{\epsilon^n}\,,
	\label{eq:bare_to_ren}
\end{equation}
	where $\rho_k=1/2$ for the gauge and Yukawa constants, $\rho_k=1$ for the scalar quartic coupling constant, and $\rho_k=0$ for the gauge fixing parameters.
	The bare couplings are defined within the dimensionally regularized~\cite{'tHooft:1972fi} theory
	with $D=4-2 \eps$.
The four-dimensional beta-functions, denoted by $\beta_i$, are defined via\footnote{For the Yukawa couplings
it is also convenient to consider the running of the coupling itself. It is obvious that
$\beta_{y_i} = (\beta_i/a_i)\,y_i/2$ with $i = t,b,\tau$.}

\begin{equation}
		\beta_i(a_k) = \frac{d a_i(\mu,\epsilon)}{d \ln \mu^2}\bigg|_{\epsilon=0},\ 		
		\qquad \beta_i = \beta_i^{(1)} + \beta_i^{(2)} + \beta_i^{(3)} + \ldots
		\label{eq:beta_def}
\end{equation}
	with $\beta^{(l)}_i$ being the $l$-loop contribution to the beta-function for $a_i$.
	The expression for $\beta_i$  can be extracted from the corresponding renormalization constants~\eqref{eq:bare_to_ren} with
	the help of
\begin{equation}
 	\beta_i =  \sum_{l}\rho_l a_l\frac{\partial c_i^{(1)}}{\partial a_l}-\rho_i c_i^{(1)}\,.
	 	\label{eq:beta_calc_simple}
	\end{equation}
	Here, again, $a_i$ stands for both the gauge couplings and the gauge-fixing.

It should be noted that the divergent part of the considered three-point functions should resemble the tree-level structure
\begin{equation}
	\Delta \mathcal{L} = - \frac{y_f}{\sqrt 2} \bar f f h - i \frac{y_f}{\sqrt 2} \bar f \gamma_5 f \chi.
	\label{eq:yukawa_verts}
\end{equation}
Since we separately consider the contributions to the 
	$\bar f_L f_R \phi$ and $\bar f_R f_L \phi$ vertices ($\phi=h,\chi$), 
	we have to be sure that the corresponding divergencies sum up to give	
	the unit matrix in the case of the Higgs boson or $\gamma_5$ in the case of $\chi$.
This serves as an additional self-consistency check of our result.

It turns out that due to the gauge symmetry the Higgs field $h$ and the would-be goldstone boson $\chi$
renormalize in the same way so that $Z_\chi = Z_h$.
Moreover, the same reasoning can be applied to the considered Yukawa vertices giving
	$Z_{ffh} = Z_{ff\chi}$.
This fact was also checked by explicit calculation at the three-loop level.

\begin{figure}[t]
	\centering
	\begin{subfigure}[b]{0.3\textwidth}
	\centering	
	\includegraphics[width=\textwidth]{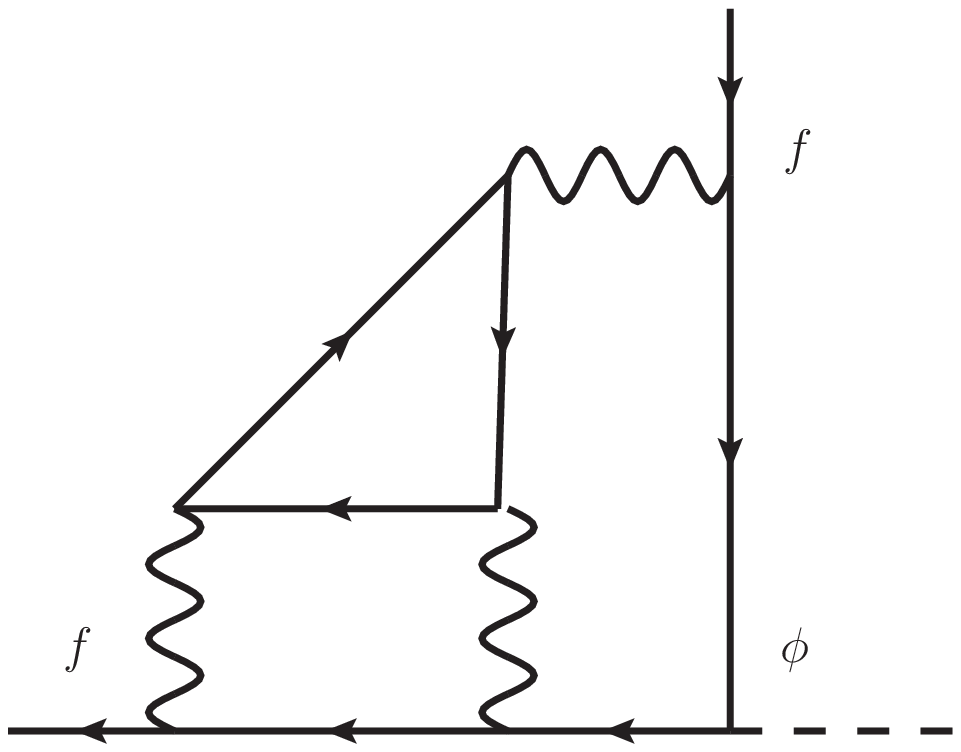}
	\caption{}
	\label{fig:a}
	\end{subfigure}
	\begin{subfigure}[b]{0.3\textwidth}
	\centering	
	\includegraphics[width=\textwidth]{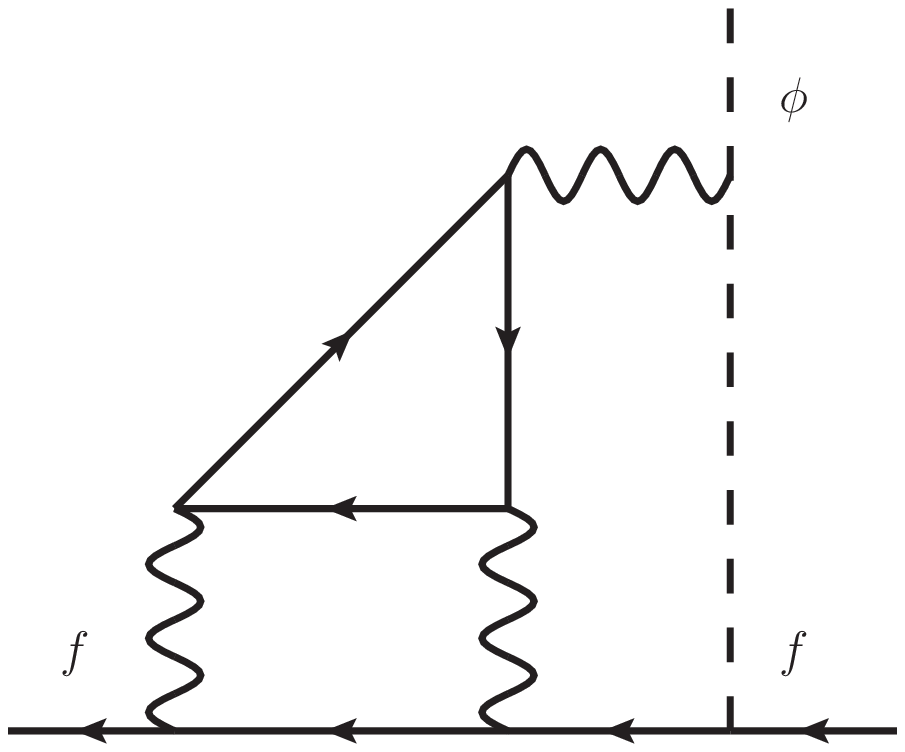}
	\caption{}
	\label{fig:b}
	\end{subfigure}
	\begin{subfigure}[b]{0.3\textwidth}
	\centering	
	\includegraphics[width=\textwidth]{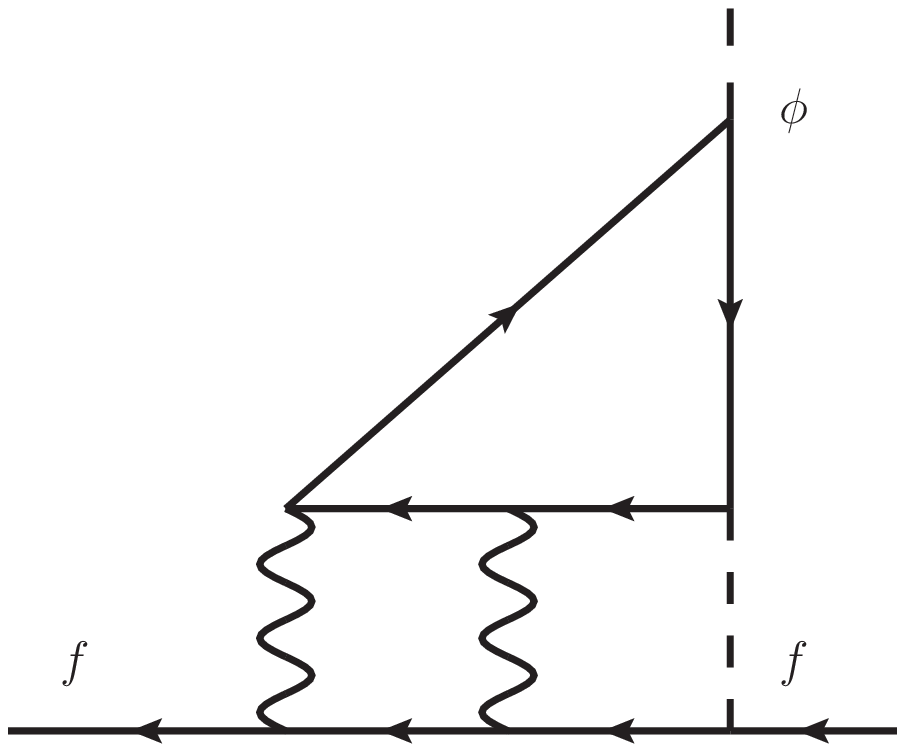}
	\caption{}
	\label{fig:c}
\end{subfigure}
	\caption{Sample Feynman diagrams giving rise to a non-trivial contribution due to the fermion traces
	with an odd number of $\gamma_5$. The second trace arises due to the projector applied to the external fermion line.
	Solid, dashed and wavy lines correspond to fermions, scalar
	and gauge bosons. In our calculation, we consider vertices with $f=t,b,\tau$ and $\phi=h,\chi$.
	Due to gauge anomaly cancellation the contributions from the first two diagrams
	are summed up to give zero if one takes into account all the SM fermions circulating
	in the triangle loops. As a consequence, we are left with the third topology which produces
	only single poles in the regularization parameter $\eps$.}
	\label{fig:eps_contrib}
\end{figure}

Actually, it is not trivial to satisfy these two requirements (to conserve the chiral structure
	of the Lagrangian and do not break the gauge invariance) at three loops.
Both the issues are related to the $\gamma_5$ problem present in dimensionally regularized theories.
It is known from the literature (see, e.g., Ref.~\cite{Jegerlehner:2000dz} and recent explicit 
	calculation~\cite{Chetyrkin:2012rz}) that the traces with an odd number of $\gamma_5$ appearing for the first time
	in the three-loop diagrams require special treatment.
We closely follow the semi-naive approach presented in Ref.~\cite{Mihaila:2012pz}.
First of all, we anticommute $\gamma_5$ with other matrices and use $\gamma_5^2 = 1$.
In the ``even'' traces all $\gamma_5$ are contracted with each other, so the corresponding expressions can
	be treated naively in dimensional regularization.
In ``odd'' traces we are left with only one $\gamma_5$.
These traces are evaluated as in four dimensions and produce totally antisymmetric tensors via the relation
\begin{eqnarray}
	\tr\left( \gamma^\mu \gamma^\nu \gamma^\rho \gamma^\sigma \gamma_5 \right) = -4 i \epsilon^{\mu\nu\rho\sigma}
	\label{eq:gamma5_trace}
\end{eqnarray}
	with $\epsilon^{0123} = 1$.

Since we are using both the $\gamma_5$ anticommutativity and 
	the four-dimensional relation \eqref{eq:gamma5_trace}, the cyclicity of the trace should be relinquished
	\cite{Korner:1991sx}.  
Due to this, a comment is in order about our positioning of $\gamma_5$ within the ``odd'' traces. 
It is known that different prescriptions give rise to an 
	ambiguity of order $D-4$. 
In our calculation the kinematical structure of a diagram, including the starting point 
	for a closed fermion loop, is fixed automatically by the \texttt{FeynArts} package \cite{Hahn:2000kx}, which is used for
	diagram generation.
This fact allows us to treat closed and open fermion chains along the same lines 
	and move $\gamma_5$ to the rightmost position (cf. with Ref.~\cite{Tsai:2009it}).
Moreover, one can be sure that for all the diagrams which have the same kinematical (\texttt{Generic} in the language of \texttt{FeynArts}) 
	structure the same prescription is used.

It is obvious that the $\eps$-tensors from \eqref{eq:gamma5_trace} can give a nontrivial contribution only if they are contracted
	with each other.
In a fully consistent calculation, this contraction should be carried out
	only after the renormalization, when one can safely take the limit $\eps \to 0$.
However, one can try to use the four-dimensional formula
\begin{equation}
	\epsilon^{\mu\nu\rho\sigma} \epsilon_{\alpha\beta\gamma\delta}
	= - {\mathcal T}{}^{[\mu\nu\rho\sigma]}_{[\alpha\beta\gamma\delta]},
	\qquad
	{\mathcal T}{}^{\mu\nu\rho\sigma}_{\alpha\beta\gamma\delta}=
				\delta^\mu_\alpha
				\delta^\nu_\beta
				\delta^\rho_\gamma
				\delta^\sigma_\delta,
	\label{eq:eps_contraction}
\end{equation}
	to contract two $\eps$-tensors originating from different fermion traces \emph{before} 
	the renormalization.
In~\eqref{eq:eps_contraction} the square brackets denote complete antisymmetrization 
	and $\delta^\mu_\alpha$ is the Kronecker delta. 
Due to the fact that the metric tensors in the left-hand side  of Eq.~\eqref{eq:eps_contraction}
	are treated as $D$-dimensional ones, in this case the result of contraction
	can differ from the correct one only via the terms proportional to $D-4$.
If the considered diagrams produce only single poles in $\eps$, then
	the above-mentioned difference contributes only to the finite terms
	which we can neglect in our calculation\footnote{This is due to the fact that such a complication arises only in the three-loop diagrams.}.
The same reasoning can be used to prove that the ambiguity due to the positioning of $\gamma_5$ in the ``odd'' traces
	does not affect the infinite part of such diagrams. 
This kind of approach (semi-naive treatment of $\gamma_5$) was used in Refs.~\cite{Mihaila:2012pz} and~\cite{Chetyrkin:2012rz}.
However, if a diagram gives rise to higher poles in $\eps$, then
	the formal manipulations with $\gamma_5$ and antisymmetric tensors affect also
	lower poles and, in general, lead to a wrong result
	unless special (finite) counter-terms are introduced (see, e.g.,~\cite{Larin:1993tq}).
In order to figure out whether our semi-naive treatment is sufficient for our calculation,
	we separately evaluate contributions arising from the contraction~\eqref{eq:eps_contraction}.
It turns out that some of the three-loop diagrams, both the fermion self-energies and vertices,
	give rise to the dangerous high-order poles in $\eps$ (see~ Figs.~\ref{fig:a} and~\ref{fig:b}).
A careful investigation shows that all these diagrams have triangle fermion (sub)loops with external gauge
	fields.
It is known that in the SM this kind of contributions should cancel if one takes into account all the
	fermion species circulating in the loops~\cite{Bouchiat:1972iq,Gross:1972pv}. 
Indeed, taking into account that the fermion traces in different diagrams contributing to Figs.~\ref{fig:a} and~\ref{fig:b} 
	are treated along the same lines and that there are no gauge anomalies in the SM, 
	one can check that all these terms sum up to zero.
It is worth mentioning that in the SM we have to put the number of colors to be equal to three ($\NR=3$)
	to satisfy the anomaly cancellation conditions.
In the end, we are left with diagrams presented in Fig.~\ref{fig:c}, which produce only single poles and give a nontrivial contribution to the final result
	(this kind of diagrams was also discussed in Ref.~\cite{Chetyrkin:2012rz}).

As in our previous calculations we use \texttt{LanHEP}~\cite{Semenov:2010qt} and \texttt{FeynArts}~\cite{Hahn:2000kx} to generate relevant Feynman diagrams
	and the \texttt{FORM}~\cite{Vermaseren:2000nd} packages \texttt{MINCER}~\cite{Gorishnii:1989gt}
	and \texttt{COLOR}~\cite{vanRitbergen:1998pn} to obtain the resulting expression. 	
Since \texttt{MINCER} is aimed only at self-energy type diagrams, 
	we made use of the infrared rearrangement~(IRR)
	trick~\cite{Vladimirov:1979zm} to evaluate the three-point Green functions by setting
	the momentum, which enters through the higgs external line, to zero\footnote{A similar method was successfully used in our previous calculations~\cite{Velizhanin:2008rw,Velizhanin:2010vw}.}.
A special script was written that automatically maps the \texttt{Feynarts} three-point vertices with zero external
	momentum to the \texttt{MINCER} topologies.

As a result of our calculation we obtain the expressions\footnote{All the results, including necessary renormalization constants, can be found online as ancillary files of the \texttt{arXiv} version of the paper.}
for the three-loop Yukawa coupling beta-functions of the heavy SM fermions. Due to lack of space we present here only the three-loop result for the
top quark Yukawa coupling ($\lam\equiv a_\lambda$):
{\allowdisplaybreaks
\begin{align}
\frac{\beta^{(1)}_t}{a_t} & = -8 \alc +\frac{9 \alu}{2}
+\ale
+\frac{3 \ald}{2}
-\frac{9 \alb}{4}
-\frac{17 \ala}{20},
\label{eq:betayu_1}
\\
\frac{\beta^{(2)}_t}{a_t} & =
\alc^2 \Big(\frac{80 \NGen}{9}-\frac{404}{3}\Big)
+36 \alc \alu
+6 \lam^2
-12 \alu \lam
-12 \alu^2
+4 \alc \ald
+9 \alb \alc
+\frac{19 \ala \alc}{15}
\notag\\
&
+\alb^2 \Big(\NGen-\frac{35}{4}\Big)
+\ala^2 \Big(\frac{29 \NGen}{45}+\frac{9}{200}\Big)
-\frac{9 \ale \alu}{4}
-\frac{11 \ald \alu}{4}
+\frac{225 \alb \alu}{16}
+\frac{393 \ala \alu}{80}
\notag\\
&
-\frac{9 \ale^2}{4}
+\frac{5 \ald \ale}{4}
+\frac{15 \alb \ale}{8}
+\frac{15 \ala \ale}{8}
-\frac{\ald^2}{4}
+\frac{99 \alb \ald}{16}
+\frac{7 \ala \ald}{80}
-\frac{9 \ala \alb}{20},
\label{eq:betayu_2}
\\
\frac{\beta^{(3)}_t}{a_t} & =
\alc^3 \Big(\frac{1120 \NGen^2}{81}+\frac{640 \NGen \z3}{3}+\frac{8864 \NGen}{27}-2498\Big)
+\alc^2 \alu \Big(-54 \NGen-228 \z3+\frac{4799}{6}\Big)
\notag\\
& +16 \alc \alu \lam
 -157 \alc \alu^2
+\alu^3 \Big(\frac{27 \z3}{2}+\frac{339}{8}\Big)
-36 \lam^3
+\frac{15 \alu \lam^2}{4}
+198 \alu^2 \lam
	\notag\\
	&
+\ala \alc^2 \Big(-\frac{176 \NGen \z3}{15}+\frac{88 \NGen}{9}-\frac{127}{60}\Big)
+\alb \alc^2 \Big(-48 \NGen \z3+38 \NGen+\frac{531}{4}\Big)
\notag \\
&
+\alc^2 \ald \Big(-\frac{14 \NGen}{3}-44 \z3-\frac{277}{2}\Big)
+\ala^2 \alc \Big(-\frac{748 \NGen \z3}{75}+\frac{5281 \NGen}{900}-\frac{1187}{300}\Big)
\notag \\
&
+\alb^2 \alc \Big(-36 \NGen \z3+\frac{57 \NGen}{4}+66\Big)
+\alb \alc \alu (180 \z3-168)
+\ala \alc \alu \Big(36 \z3-\frac{126}{5}\Big)
\notag \\
&
+\ala \alc \ald \Big(-\frac{28 \z3}{5}-\frac{457}{30}\Big)
+\alc \ald \alu (27-32 \z3)
+\alc \ald^2 (82-64 \z3)
+\frac{5 \alc \ale \alu}{2}
\notag\\
&
+\alb \alc \ald \Big(-108 \z3-\frac{27}{2}\Big)
-\frac{43}{6} \alc \ald \ale
-\frac{321}{20} \ala \alb \alc
+\alb \ald^2 \Big(\frac{63 \z3}{2}-\frac{2283}{32}\Big)
\notag \\
&
+\alb^3 \Big(\frac{50 \NGen^2}{9}+45 \NGen \z3-\frac{1139 \NGen}{144}+\frac{45 \z3}{8}-\frac{14677}{576}\Big)
+\ala \alb \alu \Big(\frac{369 \z3}{20}+\frac{8097}{640}\Big)
\notag\\
&
+\ala \alb^2 \Big(-\frac{9 \NGen \z3}{5}-\frac{9 \NGen}{80}-\frac{27 \z3}{40}+\frac{927}{320}\Big)
+\alb \ald \ale \Big(9 \z3-\frac{153}{8}\Big)
+\alb \ale^2 \Big(9 \z3-\frac{315}{16}\Big)
\notag \\
&
+\ala^2 \alb \Big(-\frac{51 \NGen \z3}{25}+\frac{241 \NGen}{400}-\frac{153 \z3}{200}+\frac{3243}{1600}\Big)
+\ala \ale \alu \Big(\frac{24 \z3}{5}-\frac{63}{5}\Big)
\notag \\
&
+\ala^3 \Big(\frac{146 \NGen^2}{81}-\frac{323 \NGen \z3}{75}+\frac{53413 \NGen}{10800}-\frac{153 \z3}{1000}+\frac{18103}{24000}\Big)
+\ald^3 \Big(\frac{9 \z3}{2}+\frac{477}{16}\Big)
\notag \\
&
+\ale^3 \Big(3 \z3+\frac{71}{16}\Big)
+\ala \alb \ald \Big(\frac{27 \z3}{10}+\frac{747}{128}\Big)
+\ala \ald^2 \Big(\frac{19 \z3}{10}-\frac{959}{160}\Big)
+\ala \ald \alu \Big(\frac{\z3}{2}-\frac{1383}{160}\Big)
\notag \\
&
+\ala^2 \alu \Big(-\frac{115 \NGen}{16}-\frac{93 \z3}{200}-\frac{44179}{19200}\Big)
+\ala^2 \ald \Big(-\frac{23 \NGen}{240}-\frac{199 \z3}{200}-\frac{35153}{19200}\Big)
\notag \\
&
+\ala \alb \ale \Big(-\frac{9 \z3}{5}-\frac{1041}{320}\Big)
+\alb \ald \alu \Big(-\frac{9 \z3}{2}-\frac{2307}{32}\Big)
+\ala \ald \ale \Big(\frac{491}{120}-\frac{27 \z3}{5}\Big)
\notag \\
&
+\ala \ale^2 \Big(-\frac{27 \z3}{5}-\frac{27}{16}\Big)
+\ala^2 \ale \Big(-\frac{117 \NGen}{40}-\frac{807 \z3}{100}-\frac{4043}{640}\Big)
+\alb \ale \alu \Big(-9 \z3-\frac{81}{4}\Big)
\notag \\
&
+\alb^2 \ale \Big(-\frac{21 \NGen}{8}-\frac{81 \z3}{4}+\frac{2121}{128}\Big)
+\alb^2 \ald \Big(-\frac{69 \NGen}{16}-\frac{225 \z3}{8}+\frac{13653}{256}\Big)
\notag  \\
&
+\ald^2 \alu \Big(\frac{825}{8}-48 \z3\Big)
+\alb^2 \alu \Big(-\frac{351 \NGen}{16}-\frac{729 \z3}{8}+\frac{49239}{256}\Big)
-\frac{45 \ale \lam^2}{2}
-\frac{291 \ald \lam^2}{4}
\notag \\
&
+45 \alb \lam^2
+9 \ala \lam^2
+30 \ale \alu \lam
+93 \ald \alu \lam
-\frac{135}{2} \alb \alu \lam
-\frac{127}{10} \ala \alu \lam
\notag \\
&
+15 \ale^2 \lam
+15 \ald^2 \lam
-\frac{171 \alb^2 \lam}{16}
+\frac{117 \ala \alb \lam}{40}
-\frac{1089 \ala^2 \lam}{400}
+\frac{21 \ale \alu^2}{2}
+\frac{739 \ald \alu^2}{16}
-\frac{1593 \alb \alu^2}{16}
\notag \\
&
-\frac{2437 \ala \alu^2}{80}
+\frac{207 \ale^2 \alu}{8}
+\frac{7 \ald \ale \alu}{2}
+\frac{53 \ald \ale^2}{4}
+22 \ald^2 \ale
	\label{eq:betayu_3}
\end{align} 
}with $\NGen$ corresponding to the number of SM generations and $\z3 = \zeta(3)$.
Having in mind the anomaly cancellation conditions, the result is only valid for the SU(3) color group, so
we substitute all the color invariants by the corresponding values ($\cR = 4/3, \NR = 3, \cA=3$).

It should be noted that the expressions are free from gauge-fixing parameters $\xiG,\xiW$ and $\xiB$
	which are present in the renormalization constants for the considered two- and three-point Green functions.
The one- and two-loop corrections, given in~Eqs.~\eqref{eq:betayu_1}~and~\eqref{eq:betayu_2}, are in
	full agreement with Refs.~\cite{Machacek:1983fi,Arason:1991ic,Luo:2002ey,Mihaila:2012pz}.  
The contribution~\eqref{eq:betayu_3} coincides with the result of Ref.~\cite{Chetyrkin:2012rz}
	in the limit of vanishing couplings $\ala,\alb,\ald$, and $\ale$ (first two lines of Eq.~\eqref{eq:betayu_3}).

In order to estimate the numerical impact of the calculated corrections, we use known values of
	the parameters at the $\MZ$ scale (e.g., from Ref.~\cite{Mihaila:2012pz}) and compute the values of the considered
	three-loop contributions to the Yukawa coupling beta-functions.
In what follows, the largest contributions that account for at least 99\% of the total three-loop corrections to the beta-functions
are shown: 
\begin{align}
\frac{\beta_t^{(3)}}{\beta_t^{(3)}(\MZ)} & \simeq
	\boxed{
	        1.51 \Alu^3
	      - 0.63 \Alc \Alu^2
	      + 0.22 \Alc^2 \Alu
	       }
	       -0.11 \Alb \Alu^2
	      + \boxed{
                0.07 \Alu^2  \Lam
	        -0.06 \Alc^3}
		\ , \label{eq:betayt3_num} \\
\frac{\beta_b^{(3)}}{\beta_b^{(3)}(\MZ)} &
	\simeq
	       1.34 \Alu^3
	      -0.19 \Alc^2 \Alu
	      -0.09 \Alc^3
	      -0.06 \Alb \Alu^2
	      -0.04 \Alb \Alc \Alu
	      +0.03 \Alc \Alu^2
	      \ , \label{eq:betayb3_num}\\
\frac{\beta_{\tau}^{(3)}}{\beta_{\tau}^{(3)}(\MZ)} & \simeq
               1.19 \Alu^3
	      -0.24 \Alc \Alu^2
	      +0.09 \Alc^2 \Alu
	      -0.04 \Alb \Alu^2 
	      -0.01 \Alu^2 \Lam
	\ ,
\label{eq:betaytau3_num}
\end{align}
	where
\begin{eqnarray}
  A_i=\frac{a_i}{a_i(\MZ)}\ ,\qquad \Lambda=\frac{\lambda}{\lambda (\MZ)}\ .
\end{eqnarray}
The known results from Ref.~\cite{Chetyrkin:2012rz} are highlighted with boxes in Eq.~\eqref{eq:betayt3_num}.

It is clear that all the corrections are dominated by the terms proportional to the top Yukawa and the strong coupling constants.
However, one can see that the $\alu^2 \alb$ contribution to the top Yukawa beta-function~\eqref{eq:betayt3_num},  which was not taken into account in Ref.~\cite{Chetyrkin:2012rz},
is comparable with the $\alu^3$ and $\alu^2\lam$ terms.

To conclude, in this paper we present for the first time the expressions for three-loop Yukawa beta-functions.
The latter can be used together with the results of Refs.~\cite{Mihaila:2012fm,Bednyakov:2012rb} and~\cite{Chetyrkin:2012rz} in a precise RGE analysis not only of the SM itself but also of its extensions which introduce New Physics at a scale very much separated from the electroweak one.
However, one should keep in mind that the full three-loop beta-function for the quartic coupling
and the anomalous dimension of the Higgs mass parameter are still missing and require a dedicated study which will be
presented elsewhere.

It is also worth mentioning that the whole calculation (generation and evaluation 
of about 1 million diagrams) was carried out on a BLTP 4-core Intel i7 computer 
and took approximately 48 hours.

\subsection*{Acknowledgments}
The authors would like to thank M.~Kalmykov for stimulating discussions.
This work is partially supported by RFBR grants 11-02-01177-a, 12-02-00412-a, RSGSS-4801.2012.2.

\end{document}